\documentclass[prl,twocolumn,superscriptaddress,showpacs,nofootinbib,reprint]{revtex4-1}
 \pdfoutput=1

\usepackage{graphicx,amsmath,amssymb,bm}

\newcommand{\Trel}{T_{\rm rel}}

\begin{document}

\title{Momentum space evolution of chiral three-nucleon forces}

\author{K.\ Hebeler}
\email[E-mail:~]{hebeler.4@osu.edu}
\affiliation{Department of Physics, The Ohio State University,
Columbus, OH 43210, USA}

\begin{abstract}
A framework to evolve three-nucleon (3N) forces
in a plane-wave basis
with the Similarity Renormalization Group (SRG) is presented and applied to
consistent interactions derived from chiral effective field theory at
next-to-next-to-leading order (N$^2$LO). We demonstrate the unitarity of
the SRG transformation, show the 
decoupling of low and high momenta, and present 
the first investigation of universality in chiral 3N forces 
at low resolution scales. 
The momentum-space-evolved 3N forces are consistent and can be directly 
combined with the standard SRG-evolved two-nucleon (NN) interactions for ab-initio 
calculations of nuclear structure and reactions.
\end{abstract}


\maketitle
\begin{figure}[t!]
\includegraphics[scale=0.43,clip=]{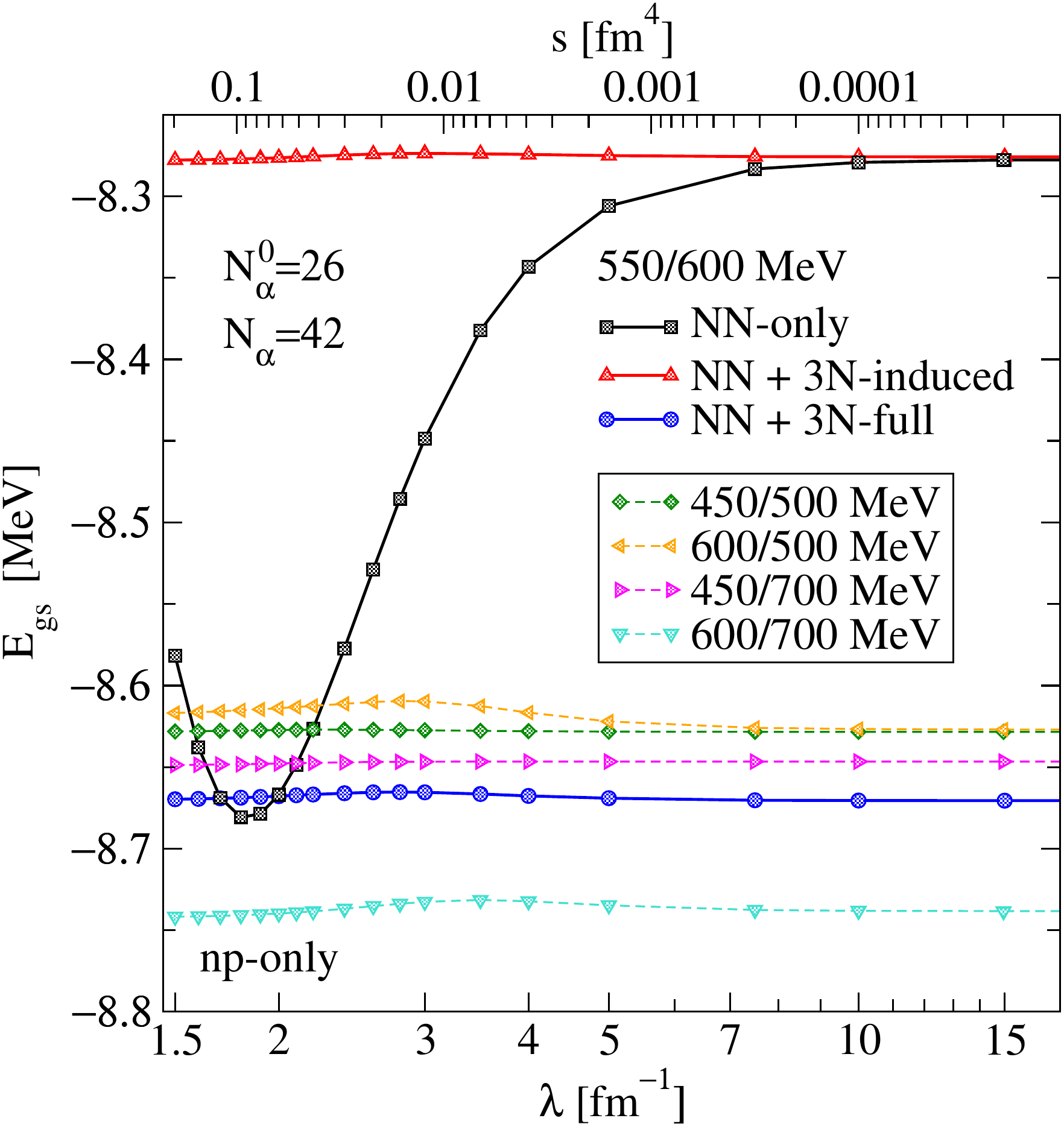}
\caption{(Color online) Ground state energy of $^3$H as a function of the flow parameter $\lambda = s^{-1/4}$ for different chiral 
interactions at N$^2$LO, which are labeled by the value of the cutoffs $\Lambda/\tilde{\Lambda}$ (see Ref.~\cite{Progpart_Epelbaum}). 
The solid lines show the results for $\Lambda/\tilde{\Lambda}=550/600$ MeV separately for NN-only, NN+3N-induced and NN+3N-full (see Ref.~\cite{Jurgenson_PRL}). The 
dashed lines show the NN+3N-full results for the other given values of $\Lambda/\tilde{\Lambda}$. Only $np$ NN interactions have been used.}
\label{fig:Triton_N2LO}
\end{figure}
\begin{figure*}[t]
\includegraphics[scale=0.95,clip=]{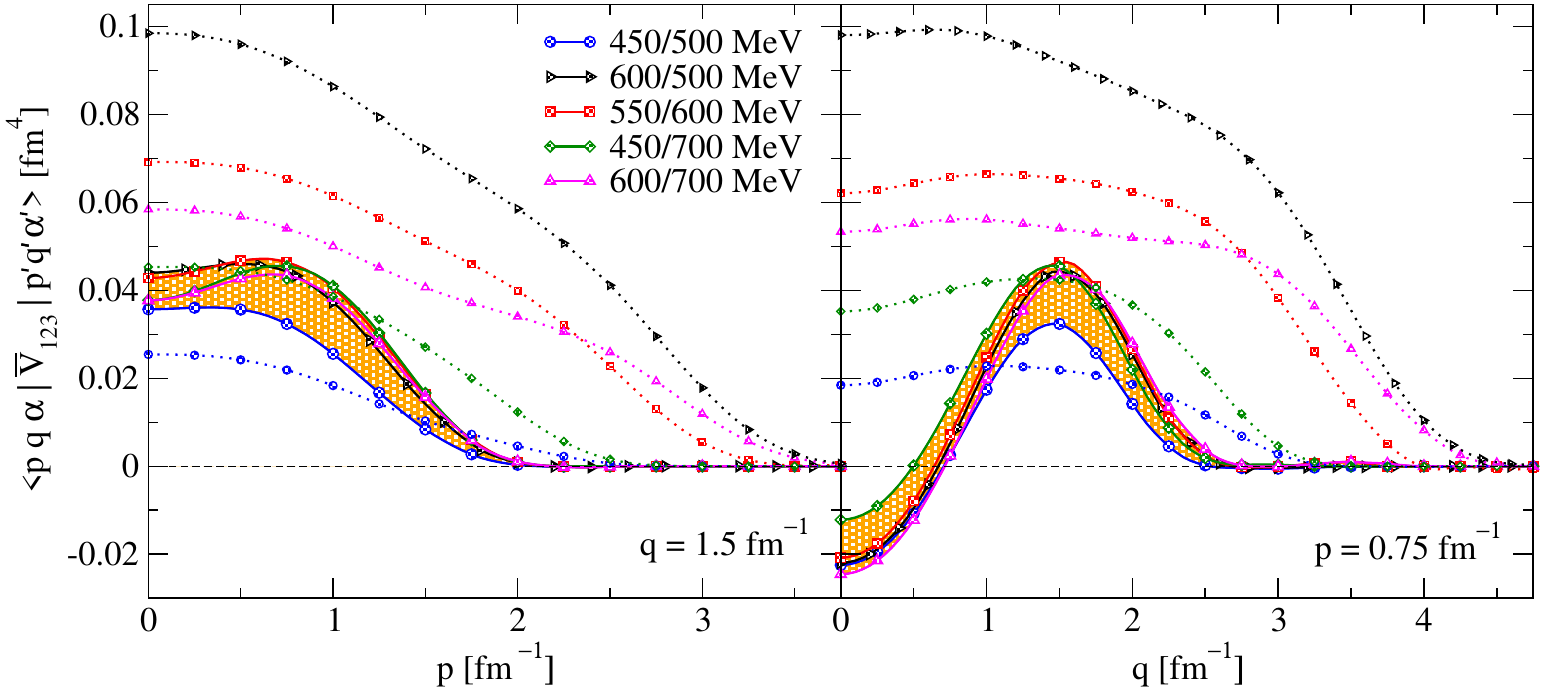}
\caption{(Color online) Matrix elements of the initial 3N forces (dotted lines) compared to evolved 3N forces (solid lines) 
at $\lambda=1.5\,\rm{fm}^{-1}$ for different interactions labeled by the values of the cutoffs $\Lambda/\tilde{\Lambda}$. 
We choose $p'=1.0\,\rm{fm}^{-1}$, $q'=1.25\,\rm{fm}^{-1}$ and $\alpha=\alpha'=2$ (see Ref.~\cite{Gloeckle_book}). 
In the left panel we fixed $q=1.5\,\rm{fm}^{-1}$ and in the right panel $p=0.75\,\rm{fm}^{-1}$. The shaded band marks 
the maximal variation between different evolved 3N force matrix elements.}
\label{fig:universal}
\end{figure*}

The interplay of chiral effective field theory (EFT) and Renormalization Group (RG) methods offers new opportunities for 
efficient and simplified microscopic calculations of many-body systems based on systematically derived nuclear forces~\cite{Epelbaum_RevModPhys,Bogner_review}. 
Recent applications range from calculations of finite nuclei~\cite{Roth_NCSM, Jurgenson_PRC, Otsuka, Holt, Bacca} 
to infinite nucleonic systems~\cite{Hebeler_PNM,Hebeler_SNM} to astrophysical applications~\cite{Hebeler_NS}.
A key challenge in all of these cases is the control of many-body forces. 
In this letter we present a new tool to address this challenge, the
RG evolution of 3N forces in a plane-wave basis (see Fig.~\ref{fig:Triton_N2LO}), which we use to 
explore universality in the evolved chiral 3N forces~(see Fig.~\ref{fig:universal}).

The SRG provides a framework to construct unitary transformations that consistently 
renormalize all operators, including many-body forces, while preserving
low-energy observables~\cite{Glazek_SRG, Wegner_SRG}. Wegner's formulation of the SRG is a
continuous series of unitary transformations of the Hamiltonian $H = \Trel + V$
as a function of the flow parameter $s$:
\begin{equation}
  \frac{d H_s}{ds} = \left[ \eta_s, H_{s} \right] \label{dHds}
  \;.
\end{equation}
Here $\eta_s$ specifies the unitary transformation, $\Trel$ denotes the relative kinetic energy and $V$ 
all the interparticle interactions. In the following we make the common choice $\eta_s = \left[ \Trel, H_{s} \right]$, 
which generates transformations that lead to a decoupling of low and high momenta in NN interactions as they are evolved
to lower resolution scales~\cite{Bogner_3N_evolution}. Such an evolution leads to much less correlated wave 
functions at low resolution and the nuclear many-body problem becomes more perturbative. 

In recent RG-based calculations, two different strategies have been 
used to handle the 3N forces.
Starting from nuclear NN and 3N forces, derived and fitted in chiral EFT, it is possible to systematically evolve
the full Hamiltonian. So far, this has been achieved by representing Eq.~(\ref{dHds}) 
using a discrete harmonic oscillator basis~\cite{Jurgenson_PRL}. 
Results for light nuclei based on such evolved interactions are promising~\cite{Jurgenson_PRC}. 
For heavier nuclei however, significant scale dependencies have been found~\cite{Roth_NCSM, Roth_CC},
which suggest that infinite matter will not be realistic. These could be indications of significant induced 
4N forces or of an insufficient evolution of 3N forces due to basis truncations.

An alternative strategy has been used for calculations of infinite nuclear systems~\cite{Hebeler_SNM}
(and elsewhere~\cite{Otsuka, Holt}). Instead of fitting the 3N forces at the chiral 
EFT cutoff scale, only the NN interactions have been evolved 
with RG methods and then the short-range parameters of the N$^2$LO 3N forces have been determined from fits to few-body systems 
at the low-momentum scale. This procedure assumes that the long-range part of the 3N forces remains invariant under the RG 
transformations and that the N$^2$LO operator structure is a 
sufficiently complete operator basis that induced contributions can
be absorbed to good approximation. The results from this procedure
are found to be in agreement with nuclear phenomenology within the theoretical uncertainties~\cite{Hebeler_PNM, Hebeler_SNM}.
However, these results do not imply that induced 3N forces have the same structure as chiral N$^2$LO 3N interactions.

A complementary evolution of nuclear many-body forces using the SRG in
a momentum plane-wave basis would allow us to reconcile
the seemingly inconsistent results of these strategies. 
In this letter, we report the first results for the evolution with Eq.~\eqref{dHds}
of realistic nuclear NN and 3N forces in such a basis. A similar approach 
has been used recently for the evolution of bosonic interaction models in one dimension~\cite{Platter}. 
As we will discuss below, however, the coupling of partial waves makes the RG evolution 
more challenging in three dimensions.

We adopt the notation and conventions of Ref.~\cite{Bogner_3N_evolution} and write 
the interaction in the form $V_s = V_{12} + V_{13} + V_{23} + V_{123}$ and the kinetic energy 
as $\Trel  = T_{12}  + T_3 = T_{13} + T_2 = T_{23} + T_1$. Here $V_{ij}$ denotes NN interactions between 
particle $i$ and $j$ and $V_{123}$ the irreducible three-body potential. 
The kinetic energy terms $T_{jk}$ and $T_i$ correspond to the contributions of the Jacobi momenta $p_i$ and   $q_i$, respectively. 
We recast Eq.~(\ref{dHds}) as separate RG flow equations for the two- and three-body 
interactions~\cite{Bogner_3N_evolution}:
\begin{eqnarray}
\frac{d V_{ij}}{ds} &=& \left[ \left[ T_{ij}, V_{ij} \right], T_{ij} + V_{ij} \right], \label{dVijds} \\
\frac{d V_{123}}{ds} &=& 
\left[ \left[ T_{12}, V_{12} \right], V_{13} + V_{23} + V_{123} \right] \nonumber \\
&&+\left[ \left[ T_{13}, V_{13} \right], V_{12} + V_{23} + V_{123} \right] \nonumber \\
&&+\left[ \left[ T_{23}, V_{23} \right], V_{12} + V_{13} + V_{123} \right] \nonumber \\
&&+\left[ \left[ \Trel, V_{123} \right], H_s \right]. \label{dV123ds}
\end{eqnarray}
Compared to Eq.~(\ref{dHds}), this system of differential equations has the important 
advantage that terms resulting from spectator particles in two-body interaction processes have 
been eliminated explicitly. In a momentum basis, the spectator particles lead to delta functions that make the
representation of Eq.~(\ref{dHds}) problematic~\cite{Gloeckle_book}.

The flow equation \eqref{dVijds} is solved in a two-body basis and then 
embedded in Eq.~(\ref{dV123ds}) by using
\begin{equation}
 V_{12} = P_{132} V_{23} P^{-1}_{132}, \quad V_{13} = P_{123} V_{23} 
 P^{-1}_{123} \, . \label{eq:embedding}
\end{equation}
Here $P_{123} = P_{12} P_{23}$ $(P_{132} = P_{13} P_{23})$ permutes three particles cyclically (anti-cyclically) with 
$P_{ij}$ denoting two-particle transpositions. We represent Eq.~(\ref{dVijds}) in a standard partial-wave basis of the form $\left| p; (L S) J T \right>$, 
where $L$, $S$, $J$ and $T$ denote the orbital angular momentum, spin, total angular momentum and isospin 
of the interacting pair of particles with relative momentum $p$. 
For the three-body basis we choose~\cite{Gloeckle_book, Stadler}
\begin{equation}
\left| p q \alpha \right> \hspace{-1.6mm} \phantom \rangle_i \equiv \left| p_i q_i; \left[ (L S) J (l s_i) j \right] \mathcal{J} \mathcal{J}_z (T t_i) 
\mathcal{T} \mathcal{T}_z \right> \, , \label{3Nmombasis}
\end{equation}
where $p_i$ and $q_i$ denote the three-body Jacobi momenta of particle $i$. The quantum numbers 
$l$, $s_i=1/2$, $j$ and $t_i=1/2$ label the orbital angular momentum, spin, total angular momentum and isospin of particle 
$i$ relative to the center-of-mass of the pair with momentum $p$. $\mathcal{J}$ and $\mathcal{T}$ denote 
the total three-body angular momentum and isospin, which are equal to $1/2$ in the three-nucleon bound states. 
For details we refer the reader to Refs.~\cite{Stadler, Gloeckle_book}.
The basis states~(\ref{3Nmombasis}) are not completely antisymmetric.
It is most natural to evolve the antisymmetrized interaction
\begin{equation}
  \bigl< p q \alpha| \overline{V}_{123} | p' q' \alpha' \bigr> \equiv  \hspace{-1mm} \phantom \langle_i \bigl< p q 
  \alpha| \mathcal{A}_{123} V_{123}^{(i)} \mathcal{A}_{123} | p' q' \alpha'   
  \bigr> \hspace{-1.2mm} \phantom \rangle_i
  \;, \label{eq:symm_int}
\end{equation}
with $\mathcal{A}_{123} = (1 + P_{123} + P_{132})$ and $V_{123}^{(i)}$ being the $i$-th Faddeev component of the three-body interaction~\cite{Gloeckle_book}.

\begin{figure}[b!]
\includegraphics[scale=0.43,clip=]{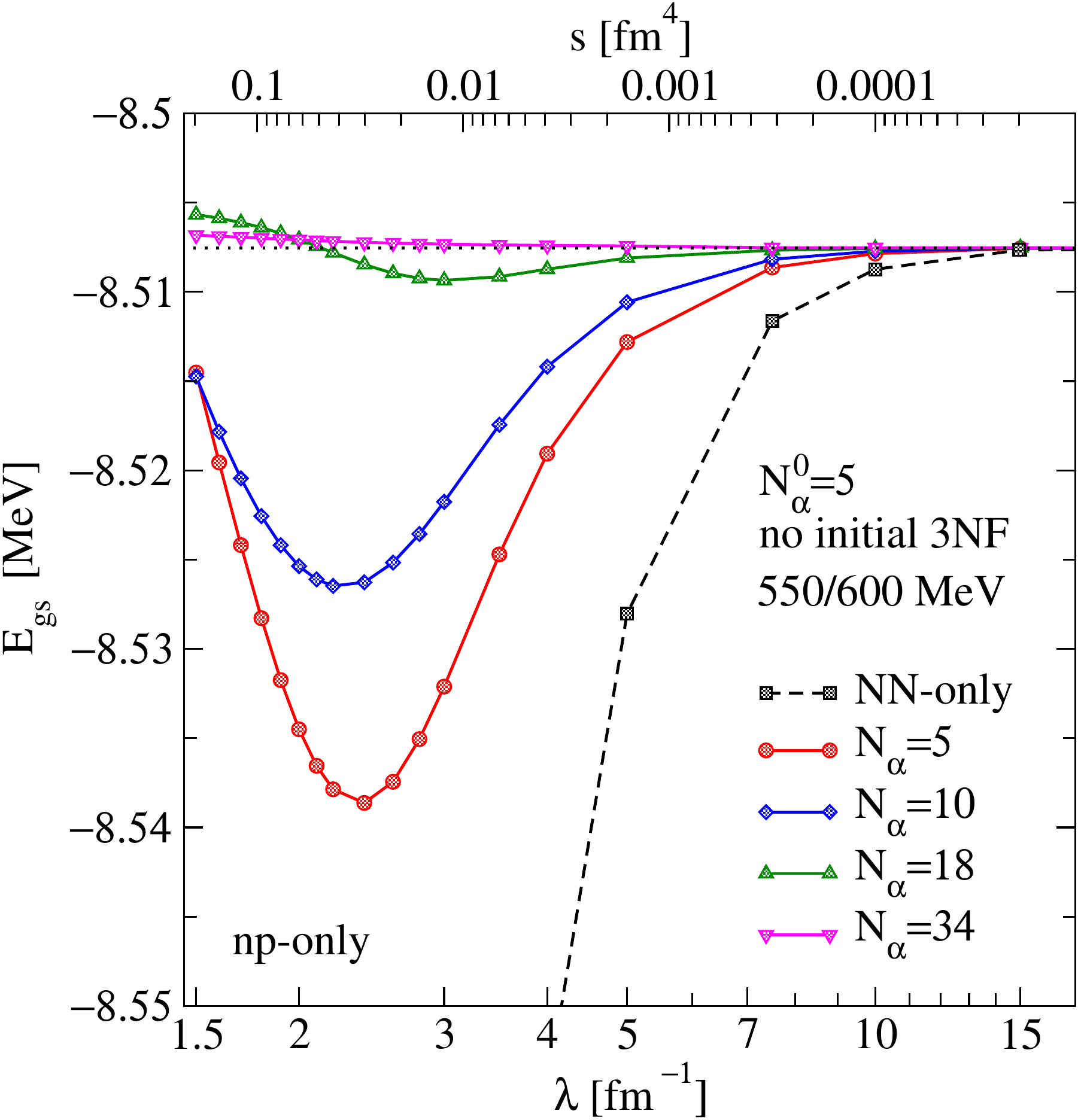}
\caption{(Color online) Ground state energy of $^3$H as a function of the flow parameter $\lambda = s^{-1/4}$ for different SRG model space 
sizes $N_{\alpha}$ and a fixed model space for the initial Hamiltonian $N_{\alpha}^0=5$ (for a detailed definition of the 3N channels see Ref.~\cite{Gloeckle_book}). The dotted line shows the energy at $\lambda = \infty$. Only $np$ NN interactions have been used.}
\label{fig:convergence}
\end{figure}

\begin{figure*}[t]
\includegraphics[scale=0.55,clip=]{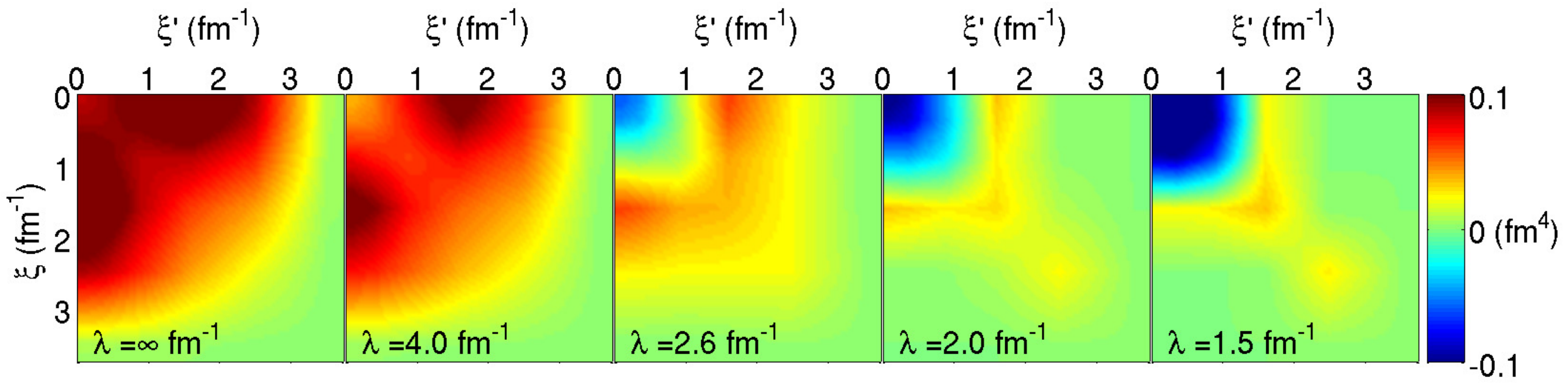} \\
\vspace{-1mm}
\includegraphics[scale=0.55,clip=]{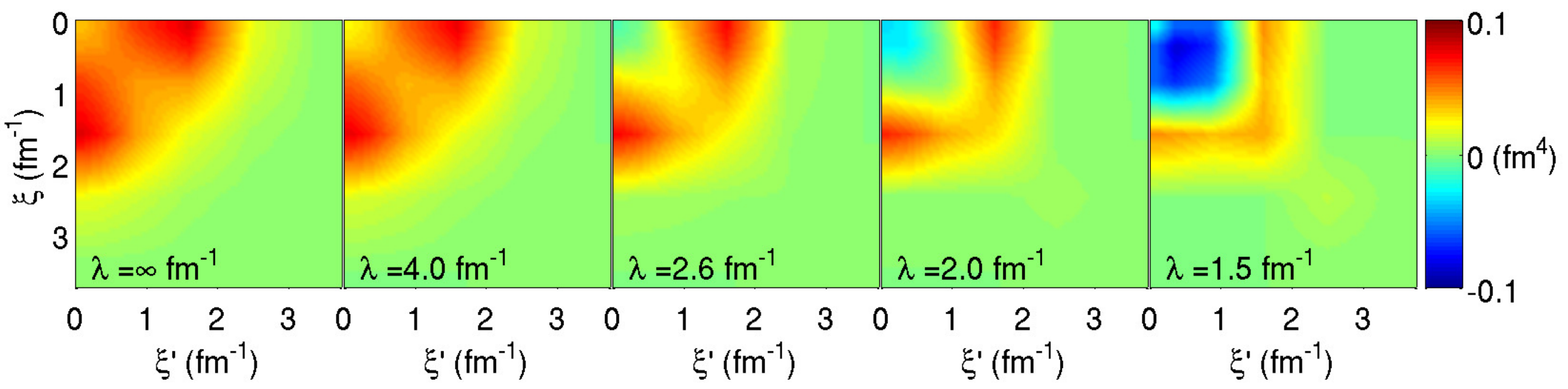}
\caption{(Color online) Contour plots of the evolved 3N potential $\left< \xi \, \alpha=1 | \overline{V}_{123} | \xi' \, \alpha'=1 \right>$ at the hyperangle $\theta = \pi/12$ (see Ref.~\cite{Gloeckle_book} for definition of partial waves). The upper panel shows the potential with $\Lambda/\tilde{\Lambda} = 550/600$ MeV and the lower panel $\Lambda/\tilde{\Lambda} = 450/500$ MeV.}
\label{fig:contour}
\end{figure*}

We solve Eqs.~(\ref{dVijds}) and (\ref{dV123ds}) for a set of five different initial NN and 3N interaction potentials derived 
from chiral EFT at next-to-next-to-leading order (N$^2$LO)~\cite{Progpart_Epelbaum}. The 3N forces low 
energy constants $c_D$ and $c_E$ have been fitted for five different cutoff combinations $\Lambda$ 
and $\tilde{\Lambda}$ to the $^3$H binding energy and the $nd$ doublet scattering length 
$^2a_{np}$~\cite{Progpart_Epelbaum,Epelbaum_private}. For this first exploratory work we represent the initial Hamiltonian 
at $s = 0$ in a basis with $N_{\alpha}^0 = 26$ three-body partial-wave channels $(J \le 3)$. The corresponding ground 
state energies are shown in Fig.~\ref{fig:Triton_N2LO}. They are found to be within 12\,keV of the converged values based 
on the antisymmetrized interaction (\ref{eq:symm_int}). We neglect the effects of charge symmetry breaking, i.e., we use 
the neutron-proton interaction in all isospin channels of the NN force. Due to this we obtain ground state energies that 
are about 150--260\,keV more bound than the experimental value $E_{\rm{gs}} = -8.481821(5)$~\cite{EB_triton} for all studied potentials.
For future applications it will be possible to incorporate the charge dependence of the nuclear interactions, as well 
the Coulomb potential, in the flow equations. 

The SRG evolution is performed in a larger model space with $N_{\alpha} = 42 > N_{\alpha}^0$ channels $(J \le 5)$. This basis size is also used for the solutions of
the Faddeev equations at the different resolution scales. As can be seen in Fig.~\ref{fig:Triton_N2LO}, the ground state energy 
is not perfectly invariant under the RG transformations. This is because a finite basis of the form~(\ref{3Nmombasis})
is not complete under cyclic and anticyclic permutations of particles, as the permuation operator
$\hspace{-1mm} \phantom \langle_i \bigl< p q \alpha | P_{ijk} | p' q' \alpha' \bigr> \hspace{-1.2mm} \phantom \langle_i$ couples in 
general all partial waves $\alpha$ and $\alpha'$~\cite{Stadler, Gloeckle_book}. 
As a consequence, non-vanishing matrix elements in all three-body partial waves are induced when two-body operators are embedded in 
a three-body momentum basis~(\ref{3Nmombasis}) via Eq.~(\ref{eq:embedding}). This problem is absent in one dimension~\cite{Platter} or in a discrete oscillator 
basis~\cite{Jurgenson_PRL}, where the permutation operator is block-diagonal in a given model space of size $N_{\rm{max}}$.
The violation of unitarity is correlated with the value of the cutoff values $\Lambda/\tilde{\Lambda}$.
For intermediate values $\Lambda/\tilde{\Lambda}=550/600$\,MeV we find a maximal variation of 5\,keV of the ground state energy, compared to about 400\,keV 
if induced 3N forces are completely neglected. For small values $\Lambda = 450$ MeV the variation is even smaller ($\le$ 2\,keV), whereas for $\Lambda=600\,$MeV it is about 16\,keV. This is natural since high momentum modes couple stronger with higher partial waves under permutation of particles. 
A signifcant part of the variations for large $\Lambda$ results from an insufficient antisymmetrization of the initial 3N forces based on a 
partial-wave representation of $\mathcal{A}_{123}$ (see Eq. (\ref{eq:symm_int})). For future applications it will be possible to optimize the evolution 
by using exactly antisymmetrized initial 3N forces (see Refs.~\cite{Golak_aPWD, Skibinski_aPWD}).

% violation of unitarity
The RG evolution can be improved systematically by increasing the model space size $N_{\alpha}$. This is demonstrated in 
Fig.~\ref{fig:convergence}. For illustrative purposes we restrict the initial Hamiltonian
at $s = 0$ to a small model space with $N_{\alpha}^0 = 5$ channels, which includes the $^1$S$_0$ and $^3$S$_1$-$^3$D$_1$ pair 
partial waves (see Ref.~\cite{Gloeckle_book}). The figure shows the triton ground state energy for different SRG model 
space sizes $N_{\alpha}$ as a function of the resolution. For the largest basis size shown, 
the resolution dependence is smaller than 1\,keV, in contrast to about 30\,keV for the smallest basis.

%universality
Low-resolution NN interactions have been found to be quantitatively very 
similar~\cite{Bogner_3N_evolution,Bogner_review}. This universality can be attributed to common long-range pion 
physics and phase-shift equivalence of all potentials, which is reflected in the matrix elements at 
low resolution. It is an interesting question if the same is true for 3N forces since there are important differences: 
First, 3N forces are fixed by fitting only two low-energy constants $c_D$ and $c_E$, in contrast to numerous couplings in
NN interactions. Second, 3N forces give only subleading contributions to observables. Since universality is only 
approximate in NN interactions, it is not obvious to what extent 3N forces are constrained by long-range physics at low resolution. 

%contour plots
In Fig.~\ref{fig:contour} we show representative examples of 3N forces matrix elements at different resolution scales. To do so we introduce 
the hyperradius $\xi^2 = p^2 + 3/4 q^2$ and the hyperangle $\tan \theta = 2p/(\sqrt{3}q)$ and choose an arbitrary 
value $\theta = \pi/12$. The upper panel shows the matrix elements of the channel $\alpha = \alpha' = 1$ for $\Lambda/\tilde{\Lambda} = 550/600$ MeV and the 
lower panel for $\Lambda/\tilde{\Lambda}=450/500$ MeV. At large resolution scales $\lambda = \infty$ ($s=0$) the off-diagonal couplings 
are much stronger for larger cutoffs $\Lambda$ and the potential is in general more repulsive. As we evolve to lower resolution, the 
repulsive off-diagonal couplings get successively suppressed and finally at $\lambda = 1.5\,\rm{fm}^{-1}$ we find matrix elements
of significant size only below $\xi, \xi' \lesssim 2\,\rm{fm}^{-1}$ and around the diagonal $\xi \simeq \xi'$. These features are
general and hold for matrix elements at all hyperangles and partial waves. In addition, the overall effects 
of the SRG evolution are stronger for initial potentials with large cutoffs $\Lambda/\tilde{\Lambda}$. In summary, the evolved 
3N matrix elements show the same tendencies found in evolved matrix elements of NN interactions~\cite{Bogner_3N_evolution}.

In Fig.~\ref{fig:universal} we show in more detail initial and low-resolution ($\lambda=1.5\,\rm{fm}^{-1}$) matrix elements for 
$\alpha=\alpha'=2$ (see Ref.~\cite{Gloeckle_book}) at some typical fixed momenta $p',q' \sim 1\,\rm{fm}^{-1}$ for all five different chiral interactions. 
We find a remarkably reduced model dependence for evolved interactions in this kinematical region. We find that this approximate collapse of matrix elements 
is most pronounced in the 3N channels in which the three-body contact interaction $c_E$ contributes ($\alpha,\alpha'= \{1,2\}$). This 
suggests that the coupling constant $c_E$ is flowing to an approximately universal value at low resolution. In addition, new momentum-dependent 
universal structures are induced at low resolution, as can be seen in the right panel. If one or more momenta becomes very small the model dependence 
of the matrix elements tends to become larger. However, the phase space of these matrix elements is suppressed and are less relevant for physical observables. 
We emphasize that all these observations are based on initial potentials that have been derived 
at N$^2$LO in chiral EFT. At this order, phase shifts at higher energies are not as well described as at N$^3$LO, which reduces the degree of 
universality of NN interactions at low resolution. It will be very interesting to investigate if universality of 3N forces becomes more pronounced by including 
contributions at N$^3$LO~\cite{Bernard_N3LO1, Bernard_N3LO2}.

The low-resolution interactions derived here are ideal for
microscopic calculations of nuclear systems.  Evolution in
a momentum basis has several advantages compared to a discrete
oscillator basis.  First, the oscillator basis has intrinsic
infrared and ultraviolet cutoffs that depend on the basis size 
and oscillator parameter $\hbar\Omega$~\cite{Jurgenson_PRC}, 
which lead to convergence issues for 3N forces.  The problems can be avoided
by first evolving in momentum space and then using a straightforward 
transformation to an oscillator basis with \emph{any} $\hbar\Omega$.
This enables new tests of SRG interactions in finite nuclei
within the no-core-shell model~\cite{Navratil_review} and 
coupled cluster~\cite{Hagen_CC3N} that allow systematic studies 
of the role of induced 4N forces. Second, the momentum-space 
interactions can be used directly
in calculations of infinite systems within many-body perturbation
theory.  This will test whether consistently evolved NN plus 3N
forces, initially fit only to few-body properties, predict empirical
nuclear saturation properties within theoretical errors, as found
previously for evolved NN forces combined with fitted 
3N forces~\cite{Hebeler_SNM}. 

Finally, since SRG transformations are usually characterized by 
the coupling of momentum eigenstates, the momentum basis 
is a natural basis in which to construct the SRG generator $\eta_s$. 
In particular, momentum-diagonal generators such as $\Trel$ (as chosen 
here) can be implemented very efficiently in a momentum basis and it 
is straightforward to generalize to the Hamiltonian-diagonal form 
advocated by Wegner~\cite{Wegner_SRG}. Other generators can significantly improve 
the efficiency of the RG evolution~\cite{Li_Generators} or achieve alternative RG decoupling
patterns, such as a flow towards a block-diagonal 
Hamiltonian~\cite{Anderson_BlockSRG}.  The possibility of using
the generator to suppress the growth of many-body forces is also under
active investigation.

\begin{acknowledgments}
I thank R.\ J.\ Furnstahl for numerous most helpful discussions, R.\ J.\ Perry and A.\ Schwenk for valuable comments, and E.\ Epelbaum 
for providing the matrix elements of the initial 3N forces. This work was supported in part by the NSF under 
Grant Nos.~PHY--0758125 and PHY--1002478, and the UNEDF SciDAC Collaboration under DOE Grant DE-FC02-07ER41457.
\end{acknowledgments}

\end{document}